\documentclass[aps,pre,preprint,showpacs,groupedaddress]{revtex4-1}
\usepackage[english]{babel}
\usepackage[T1]{fontenc}
\usepackage{natbib}
\usepackage{amssymb}
\usepackage{bm}
\usepackage{amsmath}
\usepackage[mathcal]{eucal}
\usepackage{color}
\newcommand {\nc} {\newcommand}
\nc {\beq} {\begin{eqnarray}}
\nc {\eeqn} [1] {\label{#1} \end{eqnarray}}
\nc {\eoln} [1] {\label{#1} \\}
\nc {\eol} {\nonumber \\}
\nc {\rref} [1] {(\ref{#1})}
\nc {\Eq} [1] {Eq.~(\ref{#1})}
\nc {\Ref} [1] {Ref.~\cite{#1}}
\nc {\la} {\mbox{$\langle$}}
\nc {\ra} {\mbox{$\rangle$}}
\nc {\dem} {\mbox{$\frac{1}{2}$}}
\nc {\ve} [1] {\mbox{\boldmath $#1$}}
\nc {\arrow} [2] {\mbox{$\mathop{\rightarrow}\limits_{#1 \rightarrow #2}$}}
\nc {\red}[1] {\textcolor{red}{#1}}
\begin{document}
\title{Accurate solution of the Dirac equation on Lagrange meshes}
\author{Daniel Baye}
\email[]{dbaye@ulb.ac.be}
\affiliation{Physique Quantique, C. P. 165/82, and \\
Physique Nucl\'eaire Th\'eorique et Physique Math\'ematique, C.P. 229, Universit\'e Libre de Bruxelles (ULB), B-1050 Brussels Belgium.}
\author{Livio Filippin}
\email[]{Livio.Filippin@ulb.ac.be}
\affiliation{Chimie quantique et Photophysique , C.P.\ 160/09, \\
Universit\'e Libre de Bruxelles (ULB), B-1050 Brussels, Belgium}
\author{Michel Godefroid}
\email[]{mrgodef@ulb.ac.be}
\affiliation{Chimie quantique et Photophysique , C.P.\ 160/09, \\
Universit\'e Libre de Bruxelles (ULB), B-1050 Brussels, Belgium}
\date{\today}
\begin{abstract}
The Lagrange-mesh method is an approximate variational method 
taking the form of equations on a grid because of the use of a 
Gauss quadrature approximation. 
With a basis of Lagrange functions involving associated Laguerre polynomials 
related to the Gauss quadrature, 
the method is applied to the Dirac equation. 
The potential may possess a $1/r$ singularity. 
For hydrogenic atoms, numerically exact energies and wave functions 
are obtained with small numbers $n+1$ of mesh points, 
where $n$ is the principal quantum number. 
Numerically exact mean values of powers $-2$ to 3  
of the radial coordinate $r$ can also be obtained with $n+2$ mesh points. 
For the Yukawa potential, a 15-digit agreement with benchmark 
energies of the literature is obtained with 50 mesh points or less.
\end{abstract}
\pacs{31.15.-p, 03.65.Pm, 02.70.Hm, 02.70.Jn}
\maketitle
\section{Introduction}
Numerically solving the Dirac equation raises a number of difficulties 
mostly related to the existence of the Dirac sea. 
The Dirac equation with a Coulomb potential is of particular interest 
since the existence of exact analytical results allows precise tests. 
The variational or Rayleigh-Ritz approximation for the Dirac equation 
has been discussed in depth by Grant and Quiney \cite{GQ00}. 
The authors use special spinors based on associated Laguerre polynomials. 
The B-splines variational or Galerkin method has been applied to the 
Dirac-Coulomb problem by Froese Fischer and Zatsarinny \cite{FZ09}. 
An alternative approach is the use of Bernstein B-polynomial basis sets 
\cite{BP06} that also looks promising for relativistic calculations 
of atomic properties \cite{ASP11}. 
The free-complement method also yields accurate results for this problem \cite{NN11}. 
Here we use a different numerical method, the Lagrange-mesh method, able 
to give exact energies and wave functions of this problem up to rounding errors. 
The exactness of one eigenvalue is not hindered by the much discussed problems 
of the variational collapse \cite{HK94,NN05} and 
of the kinetic balance of the basis \cite{GQ00,FZ09,NN05,STY04,Ig06,Ig07,BD08}. 

The Lagrange-mesh method is an approximate variational calculation 
using a special basis of functions, called hereafter Lagrange functions, 
related to a set of mesh points and the Gauss quadrature associated with this mesh 
\cite{BH86,VMB93}. 
It combines the high accuracy of a variational approximation and the simplicity of a calculation 
on a mesh \cite{BHV02,Ba06}. 
The Lagrange functions are $N$ infinitely differentiable functions 
that vanish at all points of this mesh, except one. 
Used as a variational basis in a quantum-mechanical calculation, 
the Lagrange functions lead to a simple algebraic system 
when matrix elements are calculated with the associated Gauss quadrature. 
The variational equations take the form of mesh equations with a diagonal representation 
of the potential only depending on values of this potential at the mesh points 
\cite{BH86,Ba06}. 
The most striking property of the Lagrange-mesh method is that, in spite of its simplicity, 
the obtained energies and wave functions can be as accurate with the Gauss quadrature 
approximation as in the original variational method 
with an exact calculation of the matrix elements \cite{BHV02,Ba06}. 
It has been applied to various problems in atomic and nuclear physics. 

Until now, most Lagrange-mesh calculations are non relativistic. 
A semi-relativistic approach based on the Salpeter equation has 
been developed in Refs.~\cite{SBH01,BS05,BS07}. 
Here we show that the Dirac equation allows a simple Lagrange-mesh treatment. 
In the case of hydrogenic atoms, it even provides numerically 
exact energies and wave functions, with very low numbers of mesh points. 
For the Yukawa potential, it can be compared with very accurate 
benchmark calculations \cite{KH02}. 

Some properties of the Dirac equation are recalled in Sec.~\ref{Dirac}. 
The Lagrange-mesh method is summarized in Sec.~\ref{LMM} 
with emphasis on its adaptation to the Coulomb-Dirac problem. 
In Sec.~\ref{Coulomb}, numerically exact energies and Dirac spinors 
are derived for hydrogenic atoms with small numbers of mesh points. 
Accurate results for the Yukawa potential are obtained and discussed in Sec.~\ref{Yukawa}. 
Sec.~\ref{conc} is devoted to concluding remarks. 

For the fine-structure constant, we use the CODATA 2010 value 
$1/\alpha = 137.035 999 074$~\cite{MTN12}. 
\section{Dirac equation for the hydrogen atom}
\label{Dirac}
In atomic units $\hbar = m_e = e = 1$ where $m_e$ is the electron mass, 
the Dirac Hamiltonian reads \cite{Gr07}
\beq
H_D = c \ve{\alpha}\cdot\ve{p} + \beta c^2 + V(r)
\eeqn{dir.1}
where $\ve{p}$ is the momentum operator, $V$ is the potential, and 
$\ve{\alpha}$ and $\beta$ are the traditional Dirac matrices. 
As the cited works use either atomic units where the speed of light $c = 1/\alpha$ 
is the inverse of the fine-structure constant, or relativistic units where $c=1$, 
we delay the full choice of units till the applications. 
The eigenenergies of $H_D$ are denoted as $c^2 + E$ 
and the Dirac equation reads  
\beq
H_D \; \phi_{\kappa m} (\ve{r}) = (c^2 + E) \;  \phi_{\kappa m} (\ve{r}).
\eeqn{dir.0}
The Dirac spinors are defined as 
\beq
\phi_{\kappa m} (\ve{r}) = \frac{1}{r} \left( \begin{array}{c} 
P_\kappa(r) \chi_{\kappa m} \\ iQ_\kappa(r) \chi_{-\kappa m} \end{array} \right)
\eeqn{dir.2}
as a function of the large and small radial components, 
$P_\kappa(r)$ and $Q_\kappa(r)$ respectively. 
The spinors $\chi_{\kappa m}$ are common eigenstates of $\ve{L}^2$, $\ve{S}^2$, 
$\ve{J}^2$, and $J_z$ with respective eigenvalues $l(l+1)$, 3/4, $j(j+1)$, and $m$ where 
\beq
j = |\kappa| + \dem, \quad l = j + \dem\, \mathrm{sgn}\, \kappa.
\eeqn{dir.3}
The coupled radial Dirac equations read in matrix form 
\beq
H_\kappa \left( \begin{array}{c} P_\kappa(r) \\ Q_\kappa(r) \end{array} \right) 
= E \left( \begin{array}{c} P_\kappa(r) \\ Q_\kappa(r) \end{array} \right)
\eeqn{dir.4}
with the Hamiltonian matrix 
\beq
H_\kappa = \left( \begin{array}{cc} 
V(r) & c \left( -\frac{d}{dr} + \frac{\kappa}{r} \right)
\\ c \left( \frac{d}{dr} + \frac{\kappa}{r} \right) & V(r) - 2c^2 \end{array} \right).
\eeqn{dir.6}
The Dirac spinors \rref{dir.2} are normed if 
\beq
\int_0^\infty \left\{ [P_\kappa(r)]^2 + [Q_\kappa(r)]^2 \right\} dr = 1.
\eeqn{dir.7}

We assume that the potential behaves at the origin as 
\beq
V(r) \arrow{r}{0} - \frac{V_0}{r}
\eeqn{dir.5}
where $V_0$ is positive or null. 
At the origin \cite{Gr07,KH02}, the radial functions behave as 
\beq
P_{\kappa}(r),\ Q_{\kappa}(r) \arrow{r}{0} r^\gamma,
\eeqn{dir.13}
with the parameter $\gamma$ defined by 
\beq
\gamma = \sqrt{\kappa^2 - (V_0/c)^2},
\eeqn{dir.12}
i.e.\ the wave functions $\phi_{\kappa m}$ are singular for $|\kappa| = 1$ 
if $V_0 \neq 0$. 
This singularity is weak for the hydrogen atom 
but can be important for hydrogenic ions with high charges $Z$ 
or for other potentials. 

An important particular case is the relativistic hydrogenic atom, 
for which the potential is 
\beq
V(r)  = -\frac{Z \alpha c}{r}
\eeqn{dir.10}
i.e.\ $V_0 = Z \alpha c$. 
As a function of the principal quantum number $n$, 
the energies are given analytically as \cite{Gr07} 
\beq
E_{n \kappa} = c^2 \left\{ \left[ 1 + \frac{\alpha^2 Z^2}{n - |\kappa| + \gamma} 
\right]^{-1/2} -1 \right\}.
\eeqn{dir.11}
They can be written in a form minimizing rounding errors as 
\beq
E_{n \kappa} = - \frac{(Z\alpha c)^2}{\mathcal{N} (\mathcal{N} + n - |\kappa| + \gamma)}
\eeqn{dir.17}
with the effective principal quantum number 
\beq
\mathcal{N} = [(n - |\kappa| + \gamma)^2 + \alpha^2 Z^2]^{1/2}.
\eeqn{dir.16}
This number is equal to $n$ when $|\kappa| = n$. 
\section{Lagrange-mesh method}
\label{LMM}
The mesh points $x_j$ are defined by \cite{BH86} 
\beq
L_N^{\alpha'}(x_j) = 0,
\eeqn{Lag.1}
where $j = 1$ to $N$ and $L_N^{\alpha'}$ is a generalized Laguerre polynomial \cite{AS65}
This mesh is associated with a Gauss quadrature 
\beq
\int_0^\infty g(x) \, dx \approx \sum^N_{k=1} \lambda_k g(x_k), 
\eeqn{Lag.4}
with the weights $\lambda_k$. 
The Gauss quadrature is exact for the Laguerre weight function $x^{\alpha'} e^{-x}$
multiplied by any polynomial of degree at most $2N-1$ \cite{Sz67}. 
The regularized Lagrange functions are defined by \cite{Ba95,BHV02,Ba06} 
\beq
\hat{f}_j (x) = \frac{x}{x_j}\, f_j(x)
= (-1)^j (h_N^{\alpha'} x_j)^{-1/2} 
\frac{L_N^{\alpha'}(x)}{x-x_j}\, x^{\alpha'/2+1} e^{-x/2}.
\eeqn{Lag.2}
In this expression, $f_j(x)$ is a standard Lagrange function \cite{BH86}. 
The functions $f_j(x)$ are polynomials of degree $N-1$ multiplied by the 
square root of the Laguerre weight $x^{\alpha'} \exp(-x)$. 
The squared norm $h_N^{\alpha'}$ of the generalized Laguerre polynomials reads 
\beq
h_N^{\alpha'} = \frac{\Gamma (N + \alpha' + 1)}{N !}.
\eeqn{Lag_2a}
The Lagrange functions satisfy the Lagrange conditions 
\beq
\hat{f}_j(x_i) = f_j(x_i) = \lambda_i^{-1/2} \delta_{ij}.
\eeqn{Lag.0}
While the explicit form of the Lagrange functions will be useful 
to choose the optimal value of $\alpha'$, 
it does not play any role in the determination of energies and mean values. 
These functions are useful when the wave functions must be known explicitly. 

The non regularized functions $f_j(x)$ form an orthonormal set 
satisfying the conditions~\rref{Lag.0} but 
have the drawback that the matrix elements of $d/dx$ and $1/x$ 
are not given accurately by the Gauss quadrature because 
the integrals contain a non polynomial factor $1/x$. 
Though the exact matrix elements are available \cite{Ba11,Ba14}, 
they lead to a variational calculation. 
The elegant simplicity of the Lagrange-mesh method is lost 
and singular potentials such as the Yukawa potential can not 
be described accurately. 
For this reason, we use in the following  the regularized functions $\hat{f}_j (x)$ for which,  
as shown below, the Gauss quadrature is exact for matrix elements of $d/dx$ and $1/x$.
This basis is however not exactly orthonormal \cite{BHV02},
\beq
\la \hat{f}_i | \hat{f}_j \ra = \delta_{ij} + \frac{(-1)^{i-j} }{\sqrt{x_ix_j}}. 
\eeqn{Lag.6}
Nevertheless, thanks to condition \rref{Lag.0}, 
these functions are orthonormal at the Gauss-quadrature approximation 
denoted with the subscript $G$, 
\beq
\la \hat{f}_i | \hat{f}_j \ra_G = \sum^N_{k=1} \lambda_k \lambda_i^{-1/2} \delta_{ik} 
\lambda_j^{-1/2} \delta_{jk} = \delta_{ij}.
\eeqn{Lag.3}
In the following, we shall treat the basis as orthonormal. 
This apparently rough approximation will be shown to have 
no effect on the physically interesting eigenvalues
and significantly simplifies the calculations. 

The matrix elements of $d/dx$ are given at the Gauss approximation by 
\beq
D_{i \neq j}^G = \lambda_i^{1/2} \hat{f}_j'(x_i) 
= (-1)^{i-j} \sqrt{\frac{x_i}{x_j}}\, \frac{1}{x_i-x_j},
\quad
D_{ii}^G = \lambda_i^{1/2} \hat{f}_i'(x_i) = \frac{1}{2x_i}.
\eeqn{Lag.20}
They are not exact since the integrands $\hat{f}_i\hat{f}'_j$ involve 
the weight function multiplied by a polynomial of degree $2N$. 
But $\int_0^\infty \hat{f}_i (\hat{f}'_j + \dem \hat{f}_j) dx$ 
can be calculated exactly with the Gauss quadrature. 
With \rref{Lag.6}, the exact expressions are thus 
\beq
D_{ij} = \la \hat{f}_i | \frac{d}{dx} | \hat{f}_j \ra 
= D_{ij}^G - \frac{(-1)^{i-j}}{2\sqrt{x_ix_j}},
\eeqn{Lag.21}
or explicitly 
\beq
D_{i \neq j} = (-1)^{i-j} \, \frac{x_i+x_j}{2\sqrt{x_ix_j}(x_i-x_j)},
\quad
D_{ii} = 0.
\eeqn{Lag.22}
This matrix is antisymmetric as expected. 

The crucial property of the Lagrange-mesh method is that the potential 
matrix elements calculated at the Gauss approximation are diagonal 
\beq
\la \hat{f}_i | V | \hat{f}_j \ra_G 
= \sum_{k=1}^N \lambda_k \hat{f}_i(x_k) V(x_k) \hat{f}_j(x_k) = V(x_i) \delta_{ij}.
\eeqn{Lag.8}
This property also applies to matrix elements of powers of $x$, for example. 
Notice that the Gauss quadrature is exact for $x^{-1}$ and $x^{-2}$ 
because the integrand is then a polynomial of degree $2N-1$ or $2N-2$ 
multiplied by the Laguerre weight function \cite{Sz67}. 

Let us now apply the method to the Dirac equation. 
To this end the radial functions $P_\kappa(r)$ and $Q_\kappa(r)$ are expanded 
in regularized Lagrange functions \rref{Lag.2} as 
\beq
P_\kappa(r) = h^{-1/2} \sum_{j=1}^{N} \; p_j \hat{f}_j^{(\alpha')}(r/h),
\eoln{Lag.10}
Q_\kappa(r) = h^{-1/2} \sum_{j=1}^{N} \; q_j \hat{f}_j^{(\alpha')}(r/h)
\eeqn{Lag.11}
where $h$ is a scaling parameter aimed at adapting the mesh points $hx_i$ 
to the physical extension of the problem. 
The superscript added to the Lagrange functions corresponds to the superscript 
of the generalized Laguerre polynomials in \Eq{Lag.2}. 

Before choosing the parameter $\alpha'$, it is important to first analyze the behavior 
of the wave functions at the origin. 
The Lagrange functions \rref{Lag.2} behave as 
\beq
\hat{f}_j^{(\alpha')}(x) \arrow{x}{0} x^{\alpha'/2 + 1}.
\eeqn{Lag.14}
Hence rather than choosing $\alpha' = 0$ like in the non-relativistic case, 
it is convenient to choose 
\beq
\alpha' = 2 ( \gamma - 1).
\eeqn{Lag.15}
If non regularized Lagrange functions were used, 
the optimal choice would be $\alpha' = 2\gamma$ 
like the one adopted in Refs.~\cite{GQ00,KH02} for the B-spline expansions. 

Let us introduce expansions \rref{Lag.10} and \rref{Lag.11} in the coupled radial 
Dirac equations \rref{dir.4}. 
A projection on the Lagrange functions leads to the $2N \times 2N$ 
algebraic system of equations 
\beq
\left( \begin{array}{cc} H^{(1,1)} & H^{(1,2)} \\ H^{(2,1)} & H^{(2,2)} \end{array} \right) 
\left( \begin{array}{c} (p_1, p_2, \dots , p_N)^T \\ 
(q_1, q_2, \dots , q_N)^T  \end{array} \right) 
= E \left( \begin{array}{c} (p_1, p_2, \dots , p_N)^T \\ 
(q_1, q_2, \dots , q_N)^T  \end{array} \right), 
\eeqn{Lag.17}
where $T$ means transposition. 
Notice that, thanks to the Gauss approximation \rref{Lag.3} 
on the scalar product of Lagrange functions, 
the energies are simply given by the eigenvalues of the Hamiltonian matrix.
According to \rref{Lag.3} and \rref{Lag.8}, 
the diagonal $N \times N$ blocks read 
\beq
H^{(1,1)}_{ij} = V(hx_i) \delta_{ij}, \quad 
H^{(2,2)}_{ij} = (V(hx_i)-2c^2) \delta_{ij}.
\eeqn{Lag.18}
For the non-diagonal blocks, the term $c\kappa/r$ 
is given exactly by the Gauss quadrature and is diagonal. 
For the matrix elements of the first derivative $d/dr$, 
several options are possible. 
One can use the exact expressions~\rref{Lag.22} 
or use the Gauss approximation 
in the spirit of the Lagrange-mesh method. 
The exact representation of $d/dr$ is antisymmetric, as it should, 
and leads to a symmetric Hamiltonian matrix. 
It is thus more instructive to exemplify the case of the Gauss quadrature 
because the matrix representation of $d/dr$ is not antisymmetric. 
One must impose the symmetry of the Hamiltonian matrix. 
Thus, the Gauss quadrature is used either in block (2,1) or in block (1,2) 
and the remaining block is constructed by symmetry. 
Choosing the Gauss quadrature in (2,1), one obtains 
\beq
H^{(2,1)}_{ij} = \frac{c}{h} \left( D_{ij}^G + \frac{\kappa}{x_i} \delta_{ij} \right),
\quad H^{(1,2)}_{ij} = H^{(2,1)}_{ji},
\eeqn{Lag.19}
where $D_{ij}^G$ is given by \rref{Lag.20}.
Choosing (1,2), one obtains 
\beq
H^{(1,2)}_{ij} = \frac{c}{h} \left(-D_{ij}^G + \frac{\kappa}{x_i} \delta_{ij} \right),
\quad H^{(2,1)}_{ij} = H^{(1,2)}_{ji},
\eeqn{Lag.19a}
which is different.
As we shall see, using the Gauss approximations leads to negligible differences 
with respect to using the exact expression. 

The norm \rref{dir.7} is calculated with the Gauss quadrature as  
\beq
\sum_{i=1}^N \left( p_i^2 + q_i^2 \right) = 1.
\eeqn{Lag.13}
Hence normed solutions of the algebraic system \rref{Lag.17} 
provide the coefficients of expansions \rref{Lag.10} and \rref{Lag.11} 
of the large and small components. 
As explained below, in the hydrogenic cases, \Eq{Lag.13} 
is numerically exact. 
\section{Hydrogenic atoms}
\label{Coulomb}
We first consider the Dirac-Coulomb problem in atomic units where $V(r) = - Z/r$. 
With $N$ mesh points, the eigenvalues and eigenvectors of the $2N \times 2N$ Hamiltonian 
matrix \rref{Lag.17} provide the relativistic energies and the coefficients 
of the expansions \rref{Lag.10} and \rref{Lag.11} of the wave functions. 
Given the block structure \rref{Lag.18} of the mesh equations, 
one expects to obtain $N$ large negative eigenvalues with an order of magnitude 
close to $-2c^2 = -37557.73008441865$. 
The remaining $N$ eigenvalues should lie much higher in the spectrum 
i.e.\ at far less negative (or positive) values. 
If the eigenvalues are ordered by increasing values, 
the $(N+1)$th eigenvalue should approximate the lowest physical energy of the chosen partial wave 
and the following ones should approximate the energies of excited states. 
With $\alpha'$ given by \rref{Lag.15} and the choice 
\beq
h = \mathcal{N}/2Z,
\eeqn{Lag.16}
the Lagrange-Laguerre expansions \rref{Lag.10} and \rref{Lag.11} are 
able to perfectly reproduce the exact eigenfunctions. 
One of these eigenvalues can even give the numerically exact result for the 
level $n\kappa$ if $N > n - |\kappa| + 1$. 
Indeed, in this case, the large and small radial functions $P_{n \kappa}$ 
and $Q_{n \kappa}$ are polynomials of degree $n - |\kappa|$ 
multiplied by $r^\gamma$ and an exponential $\exp(-Z r/\mathcal{N})$. 
Moreover, the matrix elements of the Hamiltonian between these components 
are exactly given by the Gauss-Laguerre quadrature 
{\it even if this quadrature is not exact for individual matrix elements $D_{ij}^G$}. 
\begin{table}[!t]
\caption{Eigenvalues $E_i$ of the $\kappa = -1$ Hamiltonian matrix in \Eq{Lag.19} 
for the hydrogen atom with $N=2$ and $N=3$ mesh points 
for $\alpha' = -5.325206347372990 \times 10^{-5}$ and the optimal value \rref{Lag.16} of $h$. 
Three cases are considered: Gauss approximation in block (2,1) [\Eq{Lag.19}], 
Gauss approximation in block (1,2) [\Eq{Lag.19a}], 
and exact values of the matrix elements $D_{ij}$ [\Eq{Lag.22}].} 
\begin{center}
\begin{tabular}{cccc}
\hline
 $E_i$  & Gauss (2,1) & Gauss (1,2) & $D_{ij}$ exact \\
\hline
\multicolumn{3}{l}{$1s_{1/2}$ with $N=2$ and $h=0.5$} & \\
$E_1$ & $-37563.23037066845    $ & $-37575.71144201392    $ & $-37567.70196457392    $ \\
$E_2$ & $-37559.23015764422    $ & $-37558.74482957028    $ & $-37558.75797819424    $ \\
$E_3$ & $    -0.500059907242439$ & $    -0.500006656596554$ & $    -0.500006656596554$ \\
$E_4$ & $    -0.500006656596554$ & $    11.495683364290550$ & $     3.499354548250311$ \\
\hline
\multicolumn{3}{l}{$1s_{1/2}$ with $N=3$ and $h=0.5$} & \\
$E_1$ & $-37567.74672551926    $ & $-37592.56872922842    $ & $-37576.14959978189    $ \\
$E_2$ & $-37560.38901231535    $ & $-37559.89410614127    $ & $-37559.76494157449    $ \\
$E_3$ & $-37558.55460214643    $ & $-37558.20677066509    $ & $-37558.27250631343    $ \\
$E_4$ & $    -0.500006656596554$ & $    -0.500006656596553$ & $    -0.500006656596554$ \\
$E_5$ & $    -0.258320031170988$ & $     0.132065036600383$ & $     0.070690172696772$ \\
$E_6$ & $     2.257774354082858$ & $    25.846655340028450$ & $     9.425471838916183$ \\
\hline
\multicolumn{3}{l}{$2s_{1/2}$ with $N=3$ and $h=0.9999933434699111$} & \\
$E_1$ & $-37561.22623074784    $ & $-37567.64669799570    $ & $-37563.50388400747    $ \\
$E_2$ & $-37558.80138024441    $ & $-37558.48926076874    $ & $-37558.48060877283    $ \\
$E_3$ & $-37558.03797885244    $ & $-37557.92889920434    $ & $-37557.95606920432    $ \\
$E_4$ & $    -0.739366695081362$ & $    -0.467715743773135$ & $    -0.488828609075186$ \\
$E_5$ & $    -0.260654106967512$ & $    -0.125002080189192$ & $    -0.125002080189192$ \\
$E_6$ & $    -0.125002080189193$ & $     5.466963065804257$ & $     1.363779946938871$ \\
\hline
\end{tabular}
\end{center}
\label{tab:1}
\end{table}
Let us start by testing the ground-state energy with $N=2$, 
a scaling parameter $h=0.5$, and
$\alpha' = -5.325206347372990 \times 10^{-5}$. 
The two mesh points are given by \Eq{Lag.1}, i.e., 
\beq
x_{1,2} = 2\gamma \mp \sqrt{2\gamma}.
\eeqn{Lag.29}
The four eigenvalues are displayed in Table \ref{tab:1} 
for three different ways of treating the first derivative: 
(i) Gauss approximation \rref{Lag.19} on block (2,1), 
(ii) Gauss approximation \rref{Lag.19a} on block (1,2), 
and (iii) use of the exact expression \rref{Lag.22} of $D_{ij}$ 
immediately leading to a symmetric matrix. 
In each case, one obtains two eigenvalues below $-2c^2$ as expected. 
They correspond to pseudostates in the Dirac sea. 
One of the other two eigenvalues is {\em identical} 
(with 15 digits!) in the three cases. 
However, in case (i), a spurious eigenvalue $E_3$ appears just below 
the physical eigenvalue $E_4$. 
In the other two cases, the physical eigenvalue is $E_3$. 
Anyway, this is most probably the simplest numerical calculation 
providing 15 significant figures for the ground-state energy 
of the relativistic hydrogen atom. 
At any $r$ value, 
the Lagrange-mesh functions $P_{1s}$ and $Q_{1s}$ given by \rref{Lag.10}
and  \rref{Lag.11} differ from the exact ones only by the tiny rounding errors 
on the four coefficients $p_1$, $p_2$ and $q_1$, $q_2$, 
which are the components of the eigenvector corresponding to the physical eigenvalue.
These properties remain true for all hydrogenic ions. 

The spurious eigenvalue has probably two origins. 
First, the present basis does not satisfy the property 
of kinetic balance \cite{GQ00,FZ09,STY04}. 
Second, the Gauss approximation is not exact 
at least for the overlap of Lagrange functions and introduces 
an error even when exact values of the $D_{ij}$ are used. 
The differences between the three calculations indicate 
that the spurious eigenvalue is mainly due here to the Gauss approximation. 
This is confirmed by a variational calculation using the same regularized 
Lagrange-Laguerre basis, i.e.\ a calculation with the exact matrix elements $D_{ij}$ 
and the exact overlaps $\la \hat{f}_i | \hat{f}_j \ra$ given by \Eq{Lag.6}. 
The resulting generalized eigenvalue problem provides the same exact value $E_3$ 
as in Table \ref{tab:1} and $E_4 \approx 1.1664515$. 
Since we are interested in a single eigenvalue which is exact, 
the existence of spurious eigenvalues is not a big problem. 
They can easily be detected by their instability 
when increasing the number of mesh points. 

When $N$ increases to 3, three values are below $-2c^2$ 
and the physical eigenvalue is $E_4$ in the three cases. 
Notice that while $E_4$ is almost identical, 
the other eigenvalues are quite different and meaningless. 
If one chooses $h=0.9999933434699111$ with $N=3$ 
in agreement with \Eq{Lag.16}, 
an eigenvalue becomes exactly equal to the $2s_{1/2}$ energy in the three cases 
though rounding errors may be slightly different. 
It is $E_5$ for (ii) and (iii) but it is $E_6$ for (i). 
Notice that when $h$ is rounded to 0.9999933, the physical eigenvalue 
does not change but the other ones can be significantly modified. 

Although the variational calculation with Lagrange functions 
does not present difficulties, 
it is less simple than a Lagrange-mesh calculation 
because of the non-diagonal overlap matrix of basis functions. 
The fact that the eigenvalue problem is generalized may even lead to 
additional rounding errors when $N$ is large. 
Since the simpler Lagrange-mesh method gives the same exact 
energies and wave functions, we only use in the rest of the paper 
this method with the Gauss quadrature on block (2,1). 

\begin{table}[!ht]
\caption{Regularized Lagrange-Laguerre mesh calculations of $ n \le 3$  
energies of the relativistic $Z=1$ hydrogen atom and $Z=100$ hydrogenic ion 
calculated for given $N$ and $h$ values, for the optimal value \rref{Lag.15} 
of $\alpha'$ and for $\alpha' = 0$ ($c = 137.035999074$). 
The exact energies are identical to the values obtained with $\alpha'=2\gamma-2$ 
except possibly for one or two units on the last displayed digit.}
\begin{center}
\begin{tabular}{lcrcrlcl}
\hline
$Z = 1$ & $nlj$ & $~~\kappa~~$ & $~~h=n/2Z~~$ & $N$ & \multicolumn{1}{c}{$E_{n\kappa}$ ($\alpha'=2\gamma-2$)} & $~~N~~$ & $E_{n\kappa}$ ($\alpha'=0$) \\
\hline
&$1s_{1/2}$ & $-1$ & 0.5        & 3 & $-0.500\,006\,656\,596\,554$ & 3 & $-0.500\,006\,656\,714\,711$ \\
&$2s_{1/2}$ & $-1$ & 1          & 5 & $-0.125\,002\,080\,189\,192$ & 5 & $-0.125\,002\,080\,208\,393$ \\
&$2p_{1/2}$ & $+1$ & 1          & 4 & $-0.125\,002\,080\,189\,192$ & 4 & $-0.125\,002\,080\,192\,885$ \\
&$2p_{3/2}$ & $-2$ & 1          & 4 & $-0.125\,000\,416\,028\,976$ & 4 & $-0.125\,000\,416\,029\,900$ \\
&$3s_{1/2}$ & $-1$ & 1          & 7 & $-0.055\,556\,295\,176\,422$ & 7 & $-0.055\,556\,295\,182\,736$ \\
&$3p_{1/2}$ & $+1$ & 1.5        & 5 & $-0.055\,556\,295\,176\,422$ & 5 & $-0.055\,556\,295\,195\,238$ \\
&$3p_{3/2}$ & $-2$ & 1.5        & 5 & $-0.055\,555\,802\,091\,367$ & 5 & $-0.055\,555\,802\,096\,072$ \\
&$3d_{3/2}$ & $+2$ & 1.5        & 5 & $-0.055\,555\,802\,091\,367$ & 5 & $-0.055\,555\,802\,091\,398$ \\
&$3d_{5/2}$ & $-3$ & 1.5        & 5 & $-0.055\,555\,637\,733\,815$ & 5 & $-0.055\,555\,637\,733\,829$ \\
\hline
$Z = 100$ & $nlj$ & $\kappa$ & $~~h \approx \mathcal{N}/2Z$& $N$ 
& \multicolumn{1}{c}{$E_{n\kappa}$ ($\alpha'=2\gamma-2$)} & $~~N~~$ & $E_{n\kappa}$ ($\alpha'=0$) \\
\hline
&$1s_{1/2}$ & $-1$ & 0.005      & 3 & $-5939.195\,192\,426\,652\, $ &100 &$-5932.765$ \\
&$2s_{1/2}$ & $-1$ & 0.009\,175 & 5 & $-1548.656\,111\,829\,165\, $ &100 &$-1545.707$ \\
&$2p_{1/2}$ & $+1$ & 0.009\,175 & 4 & $-1548.656\,111\,829\,167\, $ &100 &$-1548.567$ \\
&$2p_{3/2}$ & $-2$ & 0.010      & 4 & $-1294.626\,149\,195\,190\, $ &100 &$-1294.626\,143$ \\
&$3s_{1/2}$ & $-1$ & 0.013\,906 & 7 & $ -657.945\,199\,521\,658\,9$ &100 &$ -656.436$ \\
&$3p_{1/2}$ & $+1$ & 0.013\,906 & 5 & $ -657.945\,199\,521\,658\,8$ &100 &$ -657.890$ \\
&$3p_{3/2}$ & $-2$ & 0.014\,768 & 5 & $ -582.139\,046\,840\,141\,8$ &100 &$ -582.139\,036$ \\
&$3d_{3/2}$ & $+2$ & 0.014\,768 & 5 & $ -582.139\,046\,840\,141\,9$ &100 &$ -582.139\,046\,829\,$ \\
&$3d_{5/2}$ & $-3$ & 0.015      & 5 & $ -564.025\,853\,485\,845\,0$ &100 &$ -564.025\,853\,485\,675$ \\
\hline
\end{tabular}
\end{center}
\label{tab:2}
\end{table}
The energies of the $n \leq 3$ levels are displayed in Table \ref{tab:2} 
for the cases $Z = 1$ and $Z = 100$. 
The calculations are performed with small numbers $N$ of mesh points, 
i.e.\ $N = n+2$, except for $s$ states ($n>1$) where a slightly larger 
value is used to move a spurious eigenvalue to higher energies. 
With these choices, mean values of powers $r^k$ of the coordinate 
can be calculated exactly from $k = -2$ to 3 as explained below. 
The first $E_{n\kappa}$ column contains energies obtained with the optimal $\alpha'$ 
defined in \Eq{Lag.15}. 
These energies coincide with the exact ones \rref{dir.17} except possibly 
for one or two units on the last displayed digit. 
For $Z=1$, the energies are shown as obtained with $h=n/2Z$ 
but calculations with the optimal value~\rref{Lag.16} lead 
to exactly the same displayed digits 
because the difference between the $h$ values is smaller than $10^{-5}$. 
Notice that exactly degenerate energies are obtained despite the fact 
that the meshes are quite different because of different $\alpha'$ and/or $N$ values. 
As in most other applications of the Lagrange-mesh method, 
the results are not very sensitive to the precise choice of $h$. 
Nevertheless, at some higher accuracy level, multiprecision calculations aiming at more digits 
should be made with \rref{Lag.16} to provide the exact values. 

For $Z = 100$, the results are computed for the displayed truncated value of the optimal $h$ 
given by \rref{Lag.16} since the dropped digits do not affect the significant digits of the physical energies. 
The accuracy remains excellent. 
Tiny differences appear between theoretically degenerate values. 
The relative error with the non-relativistic value $h = n/2Z$ is about $10^{-10}$. 

The last column presents calculations with standard Laguerre polynomials ($\alpha' = 0$). 
For $Z = 1$, the relative difference with the fourth column is tiny 
when the same number of mesh points is kept. 
It decreases from about $2 \times 10^{-9}$ to $3 \times 10^{-13}$ when $|\kappa|$ increases. 
The singularity induced by the difference between $\gamma$ and $|\kappa|$ is weak. 
For $Z = 100$ with the same $N$, the results are very bad (not shown). 
Even with the much larger $N = 100$ value, the accuracy remains poor 
except when $|\kappa|$ is large, i.e.\ when $\alpha'$ gets closer to an integer value 
that $\alpha' = 0$ can better simulate. 
For $|\kappa| = 1$, the relative error is larger than $10^{-3}$. 
For large $Z$ values, a correct treatment of the singularity is crucial, as expected. 

\begin{table}[ht]
\caption{Regularized Lagrange-Laguerre mesh calculations of some $ n = 30$  
energies of the relativistic hydrogen atom ($Z=1$) and hydrogenic fermium ion ($Z=100$) 
for $N = 32$ and optimal parameters $\alpha'=2\gamma-2$ and $h = \mathcal{N}/2Z$. 
The displayed relative errors $\epsilon$ depend on the code implementation 
but are given for information.
Powers of ten are indicated in square brackets.}
\begin{center}
\begin{tabular}{lrccclr}
\hline
$n$ & $~\kappa~$ & $\alpha'$ & $h$ & $~~N~~$ & \multicolumn{1}{c}{$E_{n\kappa}$} & \multicolumn{1}{c}{$\epsilon$} \\
\hline
\multicolumn{7}{c}{$Z = 1$} \\
30 & $-1$  & $-5.325\,206\,347\,372\,990$[-5] & 14.999\,987\,1 & 32 & $-0.000\,555\,556\,517\,052\,700\,9$ & $  2.2[-16]$ \\
   & $+1$  &                                  &                &    & $-0.000\,555\,556\,517\,052\,702\,9$ & $  3.8[-15]$ \\
   & $-2$  & 1.999\,973\,374\,234\, 119       & 14.999\,993\,8 &    & $-0.000\,555\,556\,023\,972\,175\,9$ & $  2.0[-15]$ \\
   & $-29$ & 55.999\,998\,163\,746\,37        & 15.000\,000\,0 &    & $-0.000\,555\,555\,564\,906\,847\,1$ & $ -4.4[-16]$ \\
   & $+29$ &                                  &                &    & $-0.000\,555\,555\,564\,906\,847\,5$ & $  4.4[-16]$ \\
   & $-30$ & 57.999\,998\,224\,954\,82        & 15             &    & $-0.000\,555\,555\,563\,773\,357\,4$ & $  0       $ \\
\hline
\multicolumn{7}{c}{$Z = 100$} \\
30 & $-1$  & $-0.632\,540\,377\,608\,241\,9$ & 0.148\,463\,49 & 32 & $-5.672\,000\,589\,766\,628$ &  $ 2.0[-15]$ \\
   & $+1$  &                                 &                &    & $-5.672\,000\,589\,766\,619$ &  $ 4.4[-16]$ \\
   & $-2$  &     1.724\,237\,615\,790\,889   & 0.149\,355\,17 &    & $-5.604\,466\,953\,036\,355$ &  $ 2.4[-15]$ \\
   & $-29$ &     55.981\,634\,556\,287\,49   & 0.149\,998\,47 &    & $-5.556\,490\,981\,728\,510$ & $-2.4[-15]$ \\
   & $+29$ &                                 &                &    & $-5.556\,490\,981\,728\,514$ & $-1.8[-15]$ \\
   & $-30$ &     57.982\,246\,922\,059\,13   & 0.15           &    & $-5.556\,377\,578\,924\,101$ & $-3.0[-15]$ \\
\hline
\end{tabular}
\end{center}
\label{tab:3}
\end{table}
The high accuracy obtained in Table \ref{tab:2} is not restricted to small $n$ values. 
Some energies for $n = 30$ obtained with $N = 32$ mesh points are displayed in Table \ref{tab:3}. 
The values of $\alpha'$ and $h$ are also given. 
The last column contains the relative error $\epsilon$ with respect to the exact value \rref{dir.17}. 
This error depends on the code implementation and may vary from one calculation to another 
as well as the last one or two digits of $E_{n\kappa}$. 
Here, for low $|\kappa|$ values, a spurious eigenvalue appears below the energy given in Table \ref{tab:3}. 
In some cases, it is probably related to the problem discussed 
in Refs.~\cite{GQ00,FZ09,STY04,Ig06,Ig07,BD08}, 
i.e.\ the fact that the basis does not satisfy the kinetic-balance criterion, 
because it also occurs in the corresponding variational calculation. 
In the other cases, it disappears when the Gauss approximation is not used. 
Finally let us note the large variation of $\alpha'$ values as a function of $|\kappa|$. 
This can be avoided by using 
\beq
\alpha' = 2 ( \gamma - |\kappa|)
\eeqn{Lag.30}
rather than \rref{Lag.15}. 
The meshes are then much more similar for all $\kappa$ values. 
The correct behavior \rref{dir.13} at the origin can still be simulated 
with a corresponding increase of the number $N$ of mesh points 
depending on $n$ rather than on $n - |\kappa|$. 
The accuracy of the results does not change much with this modification. 

Tables \ref{tab:2} and \ref{tab:3} show that the present method can provide 
numerically exact energies. 
The same is true for the corresponding wave functions, as it can be realized from the calculation of the  
mean values of powers of $r$. 
With $N \ge n - |\kappa| + 3$, 
the obtained wave functions and the corresponding Gauss quadrature lead 
to the exact mean values for the operators $r^{-2}$, $r^{-1}$, $r$, $r^2$, and $r^3$ with 
\beq
\la r^k \ra_{n\kappa} =
\la \phi_{n\kappa m}| r^k |\phi_{n\kappa m} \ra = h^k \sum_{i=1}^N 
(p_{n\kappa i}^2 + q_{n\kappa i}^2) x_i^k.
\eeqn{dir.20}
Indeed, the integrand of the exact matrix element is the weight function times a polynomial 
of degree $2n-2|\kappa|+k+2$. 
The Gauss quadrature is exact for $2N-1 \ge 2n-2|\kappa|+k+2$ 
or $0 \le k \le 2(N-n+|\kappa|)-3$. 
This is thus also valid for the norm \rref{Lag.13}. 
Thanks to the regularization, the integrand contains a factor $r^{k+2}$ 
and the integral is also exact for the negative powers $k=-1$ and $-2$. 
The exact mean values of higher positive integer powers of $r$ can also be obtained 
but with increasing numbers $N$ of mesh points. 

Mean values obtained with the conditions of Table \ref{tab:2} 
for the optimal $\alpha'$ and $h$ are displayed in Table \ref{tab:4}. 
For $k = -2$, $-1$, 1 and 2, the numerical results agree with analytical expressions 
from Table 3.2 of \Ref{Gr07} or from \Ref{Sh03}. 
If the Gauss quadrature is performed on block (1,2) rather than on block (2,1), 
the mean values are closer to the exact ones for $2p_{1/2}$ and $2p_{3/2}$ 
but they are slightly less good for $1s_{1/2}$ and $2s_{1/2}$. 
\begin{table}[ht]
\caption{Lagrange-mesh calculations of the mean values $\langle (Zr)^k \rangle$ ($k = -2$ 
to 3) for the Dirac hydrogen atom with $N=3$ ($1s_{1/2}$), $N=4$ ($2p_{1/2}$ and $2p_{3/2}$), 
and $N=5$ ($2s_{1/2}$) mesh points.} 
\begin{center}
\begin{tabular}{ccccc}
\hline
 $k$  &  $1s_{1/2}$  & $2s_{1/2}$ & $2p_{1/2}$ & $2p_{3/2}$ \\
\hline
\multicolumn{5}{c}{$Z = 1$} \\
$-2$        & \hspace*{0.1cm} 2.000\,159\,766\,116\,231 \hspace*{0.1cm}  &  \hspace*{0.1cm} 0.250\,028\,292\,269\,074 \hspace*{0.1cm} &  \hspace*{0.1cm} 0.083\,342\,024\,388\,253 \hspace*{0.1cm} &  \hspace*{0.1cm} 0.083\,334\,627\,656\,595 \hspace*{0.1cm} \\
\cite{Sh03} & 2.000\,159\,766\,116\,226 &  0.250\,028\,292\,269\,074 &  0.083\,342\,024\,388\,253 &  0.083\,334\,627\,656\,577 \\
$-1$        & 1.000\,026\,626\,740\,701 &  0.250\,008\,320\,873\,086 &  0.250\,008\,320\,873\,087 &  0.250\,001\,664\,121\,470 \\
\cite{Gr07} & 1.000\,026\,626\,740\,701 &  0.250\,008\,320\,873\,086 &  0.250\,008\,320\,873\,086 &  0.250\,001\,664\,121\,445 \\
  1         & 1.499\,973\,373\,968\,263 &  5.999\,883\,511\,521\,008 &  4.999\,883\,511\,520\,941 &  4.999\,973\,374\,233\,225 \\
\cite{Gr07} & 1.499\,973\,373\,968\,263 &  5.999\,883\,511\,521\,012 &  4.999\,883\,511\,521\,012 &  4.999\,973\,374\,234\,120 \\
  2         & 2.999\,906\,809\,597\,867 & 41.998\,495\,647\,329\,15  & 29.998\,735\,280\,816\,32  & 29.999\,707\,117\,268\,71  \\
\cite{Gr07} & 2.999\,906\,809\,597\,866 & 41.998\,495\,647\,329\,22  & 29.998\,735\,280\,817\,29  & 29.999\,707\,117\,284\,25  \\
  3         & 7.499\,687\,148\,380\,748 & 329.983\,239\,243\,076\,3  & 209.987\,712\,361\,100\,8  & 209.997\,151\,055\,590\,1  \\
\hline
\multicolumn{5}{c}{$Z = 100$} \\
$-2$        & 7.960\,417\,675\,192\,373 &  1.542\,632\,708\,400\,137 &  0.454\,380\,205\,317\,436 &  0.098\,563\,843\,941\,060\,1 \\
\cite{Sh03} & 7.960\,417\,675\,192\,391 &  1.542\,632\,708\,400\,123 &  0.454\,380\,205\,317\,370 &  0.098\,563\,843\,941\,060\,0 \\
$-1$        & 1.462\,566\,036\,503\,436 &  0.398\,505\,472\,652\,605 &  0.398\,505\,472\,652\,623 &  0.268\,511\,331\,221\,178\,9 \\
\cite{Gr07} & 1.462\,566\,036\,503\,437 &  0.398\,505\,472\,652\,604 &  0.398\,505\,472\,652\,604 &  0.268\,511\,331\,221\,178\,6 \\
  1         & 1.183\,729\,811\,195\,878 &  4.675\,861\,781\,113\,669 &  3.675\,861\,781\,113\,592 &  4.724\,237\,615\,790\,892    \\
\cite{Gr07} & 1.183\,729\,811\,195\,879 &  4.675\,861\,781\,113\,673 &  3.675\,861\,781\,113\,673 &  4.724\,237\,615\,790\,889    \\
  2         & 1.993\,081\,171\,511\,766 & 26.562\,706\,733\,046\,36  & 17.293\,451\,206\,471\,90  & 27.042\,658\,666\,244\,49     \\
\cite{Gr07} & 1.993\,081\,171\,511\,771 & 26.562\,706\,733\,046\,46  & 17.293\,451\,206\,472\,39  & 27.042\,658\,666\,244\,48     \\
  3         & 4.352\,350\,770\,363\,447 &172.545\,557\,666\,531\,8   & 98.256\,482\,525\,922\,66  & 181.84\,126\,263\,455\,46     \\
\hline
\end{tabular}
\end{center}
\label{tab:4}
\end{table}

All results until now are obtained with $h$ values varying from shell to shell 
and sometimes from level to level. 
Several highly accurate eigenvalues can also be obtained simultaneously 
with a single $h$ value per partial wave or for all partial waves. 
Relative errors on the nine lowest energies are presented in Table \ref{tab:5} 
with $N = 30$ mesh points and some average scaling parameter 
depending on $\kappa$. 
At least six eigenvalues have simultaneously a relative accuracy better than $10^{-10}$
for the various partial waves. 
The worst case is $\kappa = -1$ because of a large range of binding energies 
and thus a large range of asymptotic exponential decreases 
which must be simulated with a single $h$. 
Precise results with a single value of $h$ for all partial waves 
can be obtained with larger $N$ values. 
With $N=50$ and $h=3$, the number of eigenvalues with an accuracy better than $10^{-10}$ 
rises to at least 10 in all the $|\kappa| = 1-3$ partial waves. 
With $N=100$ and $h=5.5$, this number rises to at least 25. 
\begin{table}[!ht]
\caption{Relative errors on Lagrange-mesh calculations of 
the nine lowest energies of a calculation with $N=30$ and the optimal $\alpha'$  
for the Dirac hydrogen atom with $|\kappa| = 1-3$. 
Powers of ten are indicated in square brackets.}
\begin{center}
\begin{tabular}{crrrrrr}
\hline
 $n-l-1$  &  $s_{1/2}~~$ & $p_{1/2}~~$ & $p_{3/2}~~$ & $d_{3/2}~~$ & $d_{5/2}~~$ & $f_{5/2}~~$ \\			                    
          &  $h=1.5~$    &  $h=2.5~$   &  $h=2.5~$   & 	 $h=3.5~$  & 	$h=4~$     &  $h=4.5~$  \\
\hline
  0         & $-2.7[-15]$ & $-3.0[-14]$ & $-2.1[-14]$ & $-1.8[-14]$ & $ 2.7[-15]$ & $-1.4[-14]$ \\
  1         & $-2.8[-14]$ & $-2.3[-14]$ & $-2.2[-14]$ & $-1.4[-14]$ & $-1.2[-14]$ & $-8.5[-15]$ \\
  2         & $-3.4[-13]$ & $-1.6[-14]$ & $-2.0[-14]$ & $-1.1[-14]$ & $-8.9[-15]$ & $-6.9[-15]$ \\
  3         & $-1.1[-13]$ & $-1.3[-14]$ & $-1.4[-14]$ & $-8.2[-15]$ & $-5.9[-15]$ & $-5.7[-15]$ \\
  4         & $ 2.5[-13]$ & $-9.8[-15]$ & $-6.2[-15]$ & $-4.8[-15]$ & $-7.7[-15]$ & $-4.9[-15]$ \\
  5         & $ 2.3[-12]$ & $-7.0[-15]$ & $-2.6[-15]$ & $-4.9[-15]$ & $-6.9[-15]$ & $-1.4[-15]$ \\
  6         & $ 2.5[-07]$ & $-2.7[-15]$ & $-1.8[-15]$ & $-2.0[-15]$ & $-5.9[-15]$ & $-2.6[-15]$ \\
  7         & $ 6.6[-04]$ & $ 8.9[-10]$ & $ 1.5[-10]$ & $-1.2[-15]$ & $-4.7[-15]$ & $-3.8[-15]$ \\
  8	        & $ 7.5[-02]$ & $ 7.6[-06]$ & $ 1.8[-06]$ & $ 2.3[-11]$ & $-5.7[-15]$ & $-2.3[-15]$ \\
\hline
\end{tabular}
\end{center}
\label{tab:5}
\end{table}
\section{Yukawa potential}
\label{Yukawa}
Benchmark values with a 40-digit accuracy are given in \Ref{KH02} for 
selected Yukawa potentials 
\beq
V(r) = -V_0 \frac{e^{-\lambda r}}{r}.
\eeqn{Yuk.1}
We choose some of them to test the Lagrange-mesh method in that case. 
Switching to the Yukawa potential only requires changing 
the potential values $V(hx_i)$ in the Hamiltonian matrix (see  \Eq{Lag.8}). 
The system of units is now $\hbar = m = c = 1$. 

Potential \rref{Yuk.1} has the singular behavior \rref{dir.5} at the origin. 
Parameter $\gamma$ is thus given by \Eq{dir.12} and $\alpha'$ is chosen 
according to \Eq{Lag.15}. 
The scaling parameter $h$ and the number $N$ of mesh points are adjusted 
for each potential according to the requested goals. 
Here we want to reproduce simultaneously all the energies displayed in Table 9 
of \Ref{KH02} for a given symmetry within the double precision accuracy. 
This can be achieved with $N = 40$ or 50 and an appropriate $h$ value. 

In Table \ref{tab:y1} are shown energies $c^2 + E_{n\kappa}$ 
for two cases: $\lambda = 0.01$ and $V_0 = 0.1$ 
(corresponding to $\lambda \approx 1.37$ and $V_0 \approx 13.7$ in atomic units) 
and $\lambda = 0.04$ and $V_0 = 0.7$ 
(corresponding to $\lambda \approx 5.48$ and $V_0 \approx 95.9$ in atomic units). 
\begin{table}[ht]
\caption{Regularized Lagrange-Laguerre mesh energies 
of Yukawa potentials ($c = 1$). 
Comparison with the benchmark results of \Ref{KH02} rounded at 17 digits.}
\begin{center}
\begin{tabular}{rrcc}
\hline
$n$ & $\kappa$ & $1 + E_{n\kappa}$ & \Ref{KH02} \\
\hline
& & \multicolumn{2}{c}{$\lambda = 0.01$, $V_0 = 0.1$ ($N = 40$, $h = 16$)} \\
\hline
0 & $-1$ & \hspace*{0.5cm} $0.995\,917\,081\,971\,152$ \hspace*{0.5cm} & \hspace*{0.5cm} $0.995\,917\,081\,971\,151\,89$ \hspace*{0.5cm} \\
1 &      & $0.999\,497\,559\,778\,376$ & $0.999\,497\,559\,778\,375\,46$ \\
2 &      & $0.999\,967\,446\,168\,861$ & $0.999\,967\,446\,168\,860\,68$ \\
0 & $ 1$ & $0.999\,531\,550\,432\,223$ & $0.999\,531\,550\,432\,222\,89$ \\
1 &      & $0.999\,983\,717\,932\,084$ & $0.999\,983\,717\,932\,084\,17$ \\
0 & $-2$ & $0.999\,534\,057\,514\,086$ & $0.999\,534\,057\,514\,085\,53$ \\
1 &      & $0.999\,983\,995\,560\,747$ & $0.999\,983\,995\,560\,747\,02$ \\
\hline
& & \multicolumn{2}{c}{$\lambda = 0.04$, $V_0 = 0.7$ ($N = 50$, $h = 2$)} \\
\hline
0 & $-1$ & $0.741\,201\,083\,823\,740$ & $0.741\,201\,083\,823\,739\,90$ \\
1 &      & $0.950\,294\,103\,969\,378$ & $0.950\,294\,103\,969\,378\,01$ \\
2 &      & $0.988\,794\,022\,128\,970$ & $0.988\,794\,022\,128\,970\,38$ \\
3 &      & $0.998\,408\,251\,840\,772$ &                                 \\
0 & $ 1$ & $0.950\,966\,326\,753\,638$ & $0.950\,966\,326\,753\,637\,53$ \\
1 &      & $0.989\,310\,801\,129\,036$ & $0.989\,310\,801\,129\,036\,00$ \\
2 &      & $0.998\,718\,627\,536\,472$ &                                 \\
0 & $-2$ & $0.961\,282\,015\,004\,946$ & $0.961\,282\,015\,004\,946\,09$ \\
1 &      & $0.991\,803\,837\,230\,717$ & $0.991\,803\,837\,230\,717\,12$ \\
2 &      & $0.999\,249\,454\,384\,587$ &                                 \\
\hline
\end{tabular}
\end{center}
\label{tab:y1}
\end{table}

For the first shallower potential, $h=16$ is a good compromise 
for a simultaneous treatment of the three $\kappa=-1$ lowest bound states. 
With $N=30$, the energies of these states perfectly reproduce 
the benchmark values rounded at 15 digits. 
However, the displayed results are obtained with $N=40$ 
to improve the wave functions and the mean values discussed below. 
We do not find any other bound state. 
Within the same conditions, the $\kappa = 1$ and $-2$ energies 
are also perfect. 
It should be noted that a similar quality of energies can be obtained 
with far less mesh points when each state is studied separately. 
The same ground-state energy is obtained with only 8 mesh points 
for $h = 4.5 - 5$. 
The first excited $\kappa=-1$ energy is obtained with $N=14$ and $h \approx 10$. 
The energies of the $\kappa=1$ and $-2$ levels can also be as accurate 
with less mesh points.

For the second deeper potential, the calculations are performed 
with $N = 50$ and $h = 2$. 
Here also a 15-digit accuracy is reached under these conditions. 
For the ground state, with $h = 1$, $N = 10$ would be enough 
to get the same digits. 
For $h=1.2$, $N = 12$ is enough for the first excited level. 
With $N = 50$ and $h = 2$, one observes the existence of two additional 
negative energies. 
The energy of the third excited level is obtained with the same accuracy 
as shown by a comparison with $N = 60$. 
The presence of a fifth slightly negative energy 
gives some indication of the possible existence 
of a very weakly bound fourth excited level 
but we could not reach convergence by increasing $N$ and $h$. 
For $\kappa=1$ and $-2$ also, an additional excited level 
is obtained with high accuracy under the same conditions. 

\begin{table}[!t]
\caption{Regularized Lagrange-Laguerre mesh calculation of mean values 
$\la r^k \ra$ for Yukawa potentials with $\kappa = -1$ ($c = 1$).  
Comparison with the benchmark results of \Ref{KH02} rounded at 17 digits.}
\begin{center}
\begin{tabular}{rccc}
\hline
$n$ \hspace*{0.5cm} & \multicolumn{2}{c}{$\la r^k \ra$} & \Ref{KH02} \\
\hline
& & \multicolumn{2}{c}{$\lambda = 0.01$, $V_0 = 0.1$ ($N = 40$, $h = 16$)} \\
\hline
0 \hspace*{0.5cm} & $\la r^{-1} \ra$ & \hspace*{0.5cm}$    0.099\,831\,872\,209\,1 $ \hspace*{0.5cm} & \hspace*{0.5cm} \\
  & $\la r      \ra$ & $   15.082\,434\,128\,862\,93$  & $   15.082\,434\,128\,863\,035$ \\
  & $\la r^2    \ra$ & $  304.188\,886\,493\,121\,4 $  & $  304.188\,886\,493\,124\,41 $ \\
1 \hspace*{0.5cm} & $\la r^{-1} \ra$ & $    0.022\,947\,496\,790\,515$ &                                 \\
  & $\la r      \ra$ & $   65.043\,195\,737\,250\,43$  & $   65.043\,195\,737\,250\,814$ \\
  & $\la r^2    \ra$ & $ 4980.632\,803\,277\,178\,  $  & $ 4980.632\,803\,277\,221\,3  $ \\
2 \hspace*{0.5cm} & $\la r^{-1} \ra$ & $    0.006\,923\,052\,889\,159$ &                                 \\
  & $\la r      \ra$ & $  205.370\,791\,289\,550    $  & $  205.370\,791\,289\,537\,01 $ \\
  & $\la r^2    \ra$ & $49369.953\,038\,660\,  $\,     & $49369.953\,038\,651\,105\,   $ \\
\hline									    
& & \multicolumn{2}{c}{$\lambda = 0.04$, $V_0 = 0.7$ ($N = 50$, $h = 2$)} \\
\hline
0 \hspace*{0.5cm} & $\la r^{-1} \ra$ & $    0.978\,144\,673\,350\,53 $ &                         \\
  & $\la r      \ra$ & $    1.739\,045\,717\,021\,701$ & $    1.739\,045\,717\,021\,736\,8$\\
  & $\la r^2    \ra$ & $    4.271\,937\,620\,831\,649$ & $    4.271\,937\,620\,831\,734\,4$\\
1 \hspace*{0.5cm} & $\la r^{-1} \ra$ & $    0.257\,425\,108\,303\,809$ &                                   \\
  & $\la r      \ra$ & $    7.020\,340\,332\,559\,71 $ & $    7.020\,340\,332\,559\,795\,9$\\
  & $\la r^2    \ra$ & $   59.711\,051\,926\,518\,6  $ & $   59.711\,051\,926\,519\,476\,$ \\
2 \hspace*{0.5cm} & $\la r^{-1} \ra$ & $    0.094\,765\,809\,000\,015$ &                                   \\
  & $\la r      \ra$ & $   18.075\,446\,620\,468\,82 $ & $   18.075\,446\,620\,468\,967\,$ \\
  & $\la r^2    \ra$ & $  377.461\,035\,916\,263     $ & $  377.461\,035\,916\,266\,38$\,  \\
3 \hspace*{0.5cm} & $\la r^{-1} \ra$ & $    0.037\,265\,655\,938\,1  $ &                                   \\
  & $\la r      \ra$ & $   41.739\,979\,834\,114\,5  $ &                                   \\
  & $\la r^2    \ra$ & $ 1982.037\,553\,539\,72      $ &                                   \\
\hline
\end{tabular}
\end{center}
\label{tab:y2}
\end{table}
To test the wave functions, we have computed 
the mean values of $1/r$, $r$ and $r^2$ using the conditions of Table \ref{tab:y1}.  
The corresponding results are reported in Table \ref{tab:y2}. 
The significant digits of $\la r^{-1} \ra$ are estimated 
by a comparison with $N = 60$. 
The error is of a few units on the last displayed digit. 
The other two cases can be compared with rounded results in Table 10 of \Ref{KH02}. 
For both potentials, one observes that about 14 figures are significant. 
Not only the energies but also the wave functions are highly accurate 
in these calculations. 
\section{Conclusion}
\label{conc}
For the first time, the Lagrange-mesh method is applied to the Dirac equation. 
The choice of mesh points takes precisely into account a possible singularity 
of the potential. 
A scaling parameter allows adjusting the mesh to the extension of the physical problem. 

For the exactly solvable Coulomb-Dirac problem describing hydrogenic atoms, 
numerically exact results, i.e.\ exact up to rounding errors, 
are obtained for any state and for any nuclear charge 
with very small numbers of mesh points. 
Only two points are enough to get the exact energy and wave function of the ground state. 
With a slightly larger number of points, mean values of a number of powers 
of the coordinate are also obtained exactly with the Gauss quadrature.

Tests with the Yukawa potential provide very accurate results with a 
number of mesh points for which the computation seems instantaneous. 
The approximate wave functions provide mean values of powers of the coordinate 
that are also extremely precise. 

A more stringent test of wave functions would be given by the calculation of polarizabilities. 
For the non relativistic hydrogen atom, numerically exact polarizabilities 
can be found with the Lagrange-mesh method for small numbers of mesh points \cite{Ba12}. 
Work is in progress to extend this study to the relativistic case for which  
very accurate values are available for comparison \cite{TZZ12}. 

The present method is expected to be very accurate for all properties 
of a single particle described by Dirac equations with various potentials. 
This includes taking account of the finite extension of the nucleus, 
evaluating two-photon transition probabilities 
or studying the scattering by some potential. 
An extension to two-electron atoms should also be accurate 
if treated in perimetric coordinates \cite{HB99}. 
A big challenge is to extend the method with accuracy to polyelectronic atoms 
where several Coulomb singular terms appear. 
A simultaneous regularization of several singularities is not available at present. 
A hybrid treatment may be feasible involving Lagrange functions 
but where the associated Gauss quadrature is replaced by another numerical technique 
for the computation of the matrix elements of the Coulomb repulsion between electrons. 
\begin{acknowledgments}
This text presents research results of the interuniversity attraction pole programme P7/12 
initiated by the Belgian-state Federal Services for Scientific, Technical and Cultural Affairs.
LF acknowledges the support from the FRIA.
\end{acknowledgments}
%
%
%
\end{document}